\documentclass[
reprint,
superscriptaddress,
amsmath,
amssymb,
]{revtex4-2}
\usepackage{comment}
\usepackage[utf8]{inputenc}
\usepackage[T1]{fontenc}
\usepackage{color}
\usepackage{graphicx}% Include figure files
\usepackage{dcolumn}% Align table columns on decimal point
\usepackage{bm}% bold math
\usepackage{xfrac}
\usepackage{upgreek}
\usepackage{textgreek}

\usepackage{dcolumn}% Align table columns on decimal point
\usepackage{bm}% bold math
\usepackage{braket}
\usepackage{sidecap}
\usepackage {hyperref}
\sidecaptionvpos{figure}{c}

\graphicspath{Figures}

\begin{document}

\preprint{APS/123-QED}

\title{Purcell enhanced and tunable single-photon emission at telecom wavelengths from InAs quantum dots in circular photonic crystal resonators}

% Force line breaks with \\

%In order to put both email and equal contribution use the following:
\author{
A. Barbiero$^{1\dagger\ast}$,
G. Shooter$^{1 \dagger}$,
J. Skiba-Szymanska$^{1}$,
J. Huang$^{1}$,
L. Ravi$^{2}$,
J. I. Davies$^{2}$,
B. Ramsay$^{3}$,
D. J. P. Ellis$^{3}$,
A. J. Shields$^{1}$,
T. M\"{u}ller$^{1}$,
and R. M. Stevenson
}

\affiliation{\small
Toshiba Europe Limited, 208 Cambridge Science Park, Cambridge CB4 0GZ, UK \\
$^{2}$ IQE Europe Limited, Pascal Close, St. Mellons, Cardiff, CF3 0LW, UK \\
$^{3}$ Cavendish Laboratory, University of Cambridge, Madingley Road, Cambridge, CB3 0HE, UK \\
\medskip
$^{\dagger}$These authors contributed equally to this work;\\
$^{\ast}$andrea.barbiero@toshiba.eu;
}

\medskip
%\date{\today}

\begin{abstract}
\noindent Embedding semiconductor quantum dots into bullseye resonators has significantly advanced the development of bright telecom quantum light sources for fiber-based quantum networks. To further improve the device flexibility and stability, the bullseye approach should be combined with a \textit{pin} diode structure to enable Stark tuning, deterministic charging, and enhanced coherence. In this work, we fabricate and characterize photonic structures incorporating hole gratings that efficiently support charge carrier transport while maintaining excellent optical performance. We report bright, Purcell-enhanced single-photon emission in the telecom C-band under above-band and phonon-assisted excitation. Additionally, we present electrically contacted resonators, demonstrating wide range tuneability of quantum dot transitions in the telecom O-band. These results mark significant steps toward scalable and tunable quantum light sources for real-world quantum photonic applications. 
\end{abstract}

\maketitle


%%%%%%%%%%%%%%%%%%%%%%%%%%%%%%%%%%%%%%%%%%%%%%%%%%%%%%%%%%%%%%%%%%%%%
%% INTRODUCTION
%%%%%%%%%%%%%%%%%%%%%%%%%%%%%%%%%%%%%%%%%%%%%%%%%%%%%%%%%%%%%%%%%%%%%
\section*{Introduction}
\noindent Semiconductor quantum dots (QDs) have emerged as essential components in advancing quantum communication networks, offering near ideal quantum light sources \cite{Somaschi.2016, Ding.2016, Wang.2019b, Uppu.2020, Tomm.2021} and efficient spin-photon interfaces \cite{Gao.2012, Cogan.2023, Coste.2023, Laccotripes.2024, Meng.2024}. For their integration with current fiber infrastructure and technology, operation at telecom wavelength is as essential as wavelength tuneability, required for interfacing and multiplexing with other sources of classical or quantum light over the same optical fiber. Recent progress in coupling QDs to bullseye resonators \cite{Liu.2019, Wang.2019}, also known as circular Bragg gratings (CBGs), has enabled highly efficient telecom wavelength photon sources \cite{Barbiero.2022, Nawrath.2023, Joos.2024, Holewa.2024,Kim.2025}, paving the way for scalable quantum networking applications.

% FIGURE 1
\begin{figure*}[t]
\includegraphics[width=0.67\textwidth]{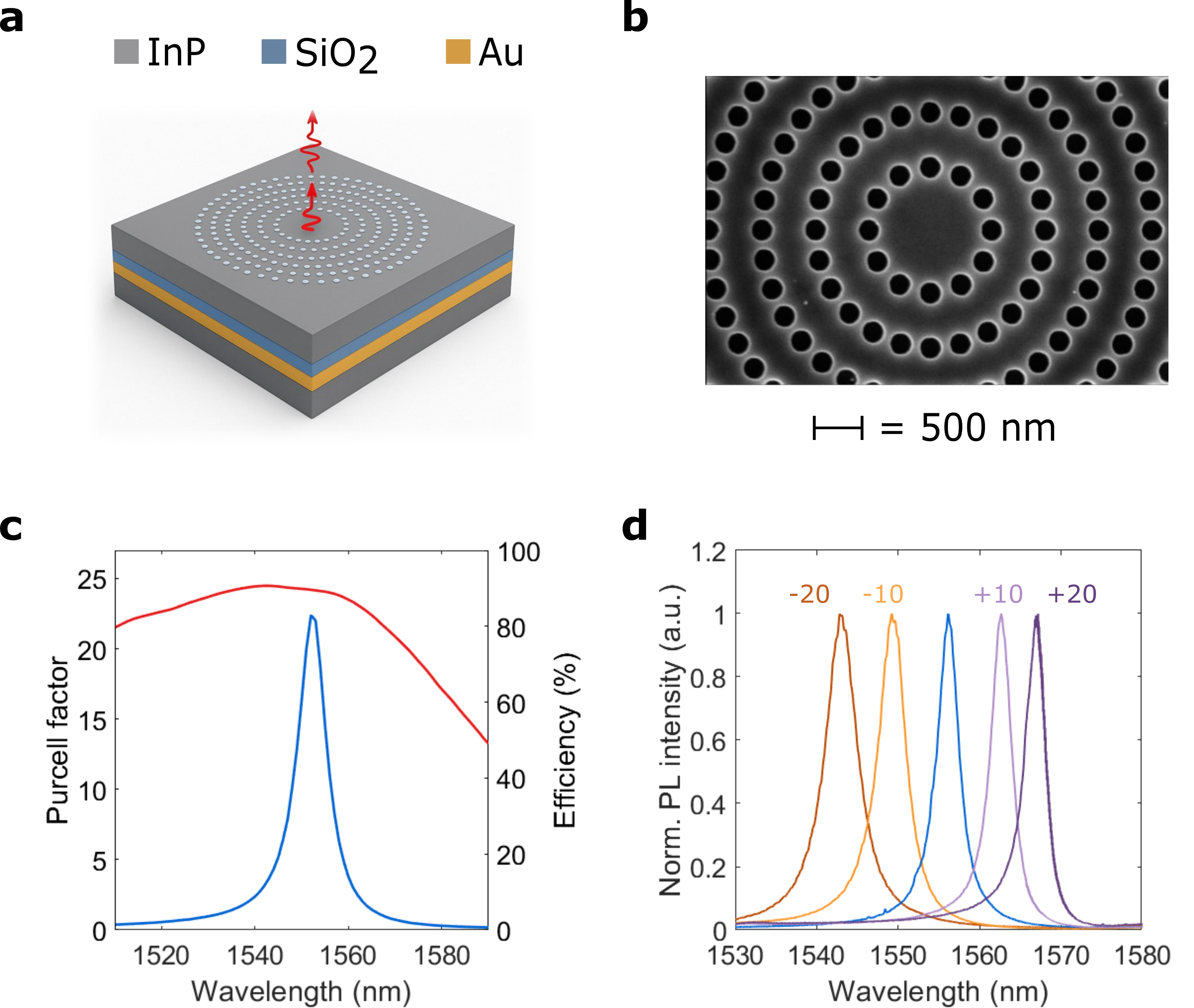}
\caption{\textbf{Circular photonic crystal resonator.} 
(a) Pictorial illustration of the device. 
(b) Scanning electron microscope top view of an exemplary InP device.
(c) Simulated performance of a device operating in the telecom C-band, with Purcell factor (blue) and dipole collection efficiency in NA = 0.65 (red) as a function of wavelength. 
(d) Cavity modes measured under high-power above-band excitation for a series of InP devices with different central disk radii. The nominal radius of the central device (blue) is set to 790 nm. Labels indicate the variation in central disk radius measured in nm.
}
\label{Fig1} 
\end{figure*}
In addition to achieving high efficiency and Purcell enhancement, precise control over the electronic states of QDs is essential for various advanced functionalities. One widely utilized technique is Stark tuning, which enables dynamic spectral control by adjusting the emission wavelength through the application of an external electric field \cite{Bennett.2010, Nowak.2014}. Other advantages of this technique are charge stabilization, which ensures highly coherent and blinking free photon emission \cite{Zhai.2020, Zhai.2022}, and deterministic charging \cite{Warburton.2000} involving the controlled introduction of charge carriers in the QD. These functionalities necessitate QDs incorporated in a a \textit{pin} diode structure, with biasing via electrical contacts on both the \textit{p}- and  \textit{n}-doped layers, ensuring effective application of electric fields. However, the fully etched trenches of conventional CBG designs isolate the central disk from the contacts, making it impossible to apply a bias directly across the QD. While early device geometries with partial etching allowed bottom contacts \cite{Ates.2012}, the implementation of a top contact remains a challenge.

% FIGURE 2
\begin{figure*}[t]
\includegraphics[width=1\textwidth]{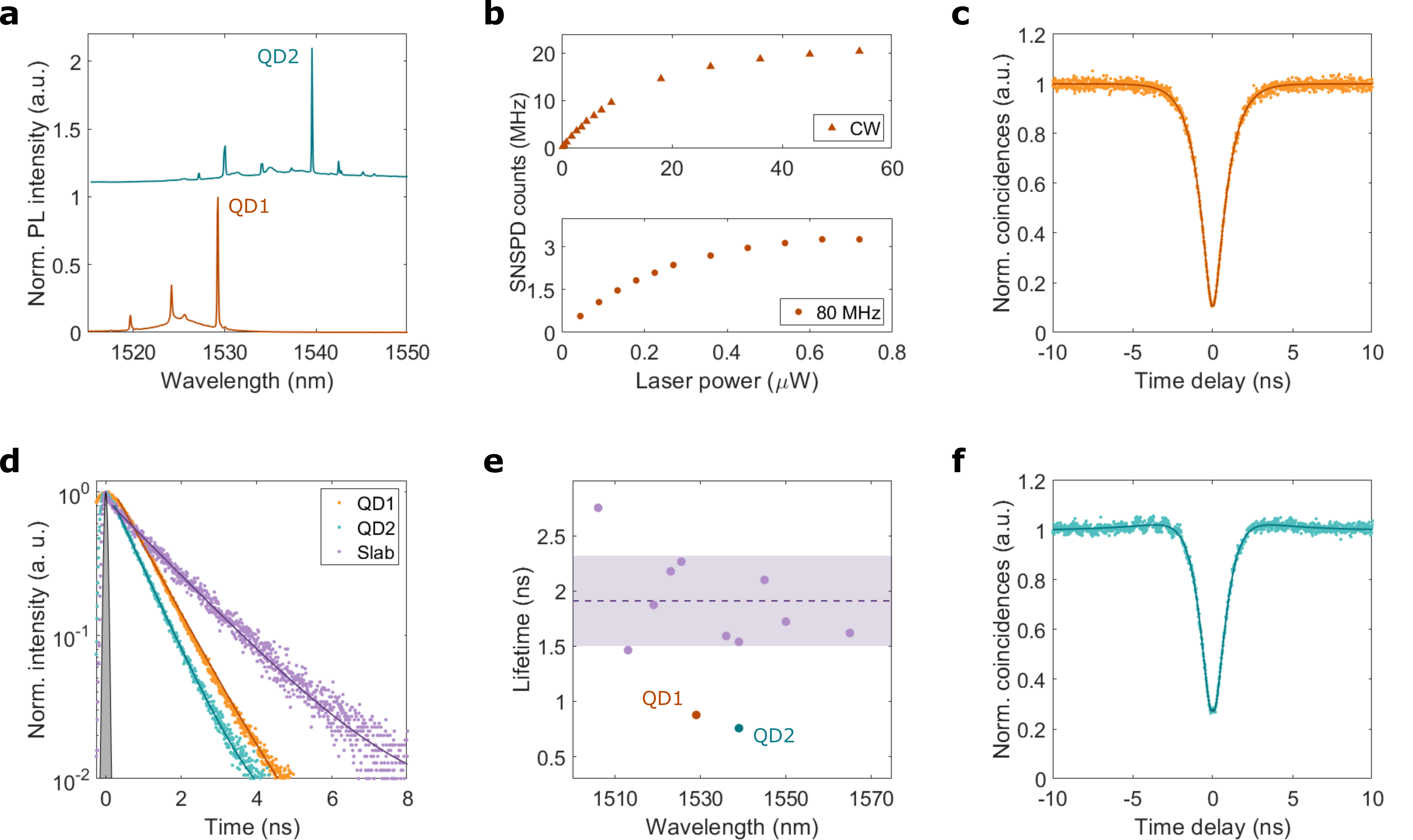}
\caption{\textbf{Device characterization under above-band excitation.} 
(a) Photoluminescence spectra of two C-band quantum dots embedded in circular photonic crystal cavities, with the brightest transitions at 1529 nm and 1540 nm labeled QD1 and QD2, respectively. 
(b) SNSPD count rates as a function of the optical pump power for QD1, measured under above-band CW and 80 MHz pulsed excitation.
(c) Second-order correlation function $g^2(\tau)$ of QD1 measured under CW excitation at half saturation power. Fitting the HBT data with a three-level model (solid line) yields $g^2(0) = 0.107(0.005)$.
(d) Radiative lifetimes of QD1 ($\sim$800 ps), QD2 ($\sim$750 ps), and of an exemplary QD transition in the bare InP slab ($\sim$1.62 ns). The gray area indicates the instrument response function, while the solid lines represent the exponential fit of the decay traces.
(e) Comparison of the radiative lifetimes of QD1 and QD2 with those of 10 transitions in the bare InP slab (purple). The dashed line and shaded region indicate the mean and standard deviation, respectively. 
(f) Second-order correlation function $g^2(\tau)$ of QD2 measured under CW excitation at half saturation power. The HBT data are fitted with a four-level model (solid line), which yields $g^2(0) = 0.281(0.006)$.
}
\label{Fig2} 
\end{figure*}

Several approaches have been proposed to address this limitation. Ridge-based CBG designs feature narrow connections between  the doped layers in the central mesa and the external metal contacts \cite{Barbiero.2022b, Singh.2022}. Though experimentally validated \cite{Wijitpatima.2024}, these devices exhibit compromises in performance compared to traditional bullseye resonators. Moreover, the presence of the ridges introduces anisotropy in the enhancement, with differing effects observed for the dipole components parallel and perpendicular to the ridge axis. This asymmetry poses challenges for advanced applications that rely on an unpolarized Purcell effect \cite{Coste.2023}. Labyrinth-like geometries \cite{Buchinger.2023}, created by rotating the connections between consecutive rings, disable waveguiding effects and improve mode confinement. On the other hand, they have only been investigated in numerical simulations and may display increased electrical resistance due to elongated paths between the central disk and external contacts. 

Recent works \cite{Jeon.2022, Ma.2024} have explored circular photonic crystal resonators, featuring air holes etched in a radially symmetric arrangement around a central disk. A design optimization targeting the telecom C-band demonstrated optical performance comparable to conventional bullseye resonators with fully etched trenches \cite{Ma.2024}. Furthermore, the gaps between the air holes were identified as potential pathways for charge carriers, enabling electric field control. Despite these promising theoretical findings, experimental validation of this design has remained elusive.
Here, we present the first experimental realization of telecom quantum light sources based on semiconductor QDs coupled to circular photonic crystal resonators. By measuring photoluminescence (PL) from InAs/InP QDs embedded in these devices, we observe bright Purcell-enhanced single-photon emission in the telecom C-band under above-band excitation. Additionally, we show that the devices are compatible with phonon-assisted excitation, which reduces multiphoton contributions and significantly improves the purity of the source. Finally, we demonstrate Stark tuning of up to 10 nm for QD transitions in the telecom O-band using electrically contacted resonators.

%%%%%%%%%%%%%%%%%%%%%%%%%%%%%%%%%%%%%%%%%%%%%%%%%%%%%%%%%%%%%%%%%%%%%
%% RESULTS
%%%%%%%%%%%%%%%%%%%%%%%%%%%%%%%%%%%%%%%%%%%%%%%%%%%%%%%%%%%%%%%%%%%%%
\section*{Results}
%
%
% FIGURE 3
\begin{figure*}[t]
\includegraphics[width=1\textwidth]{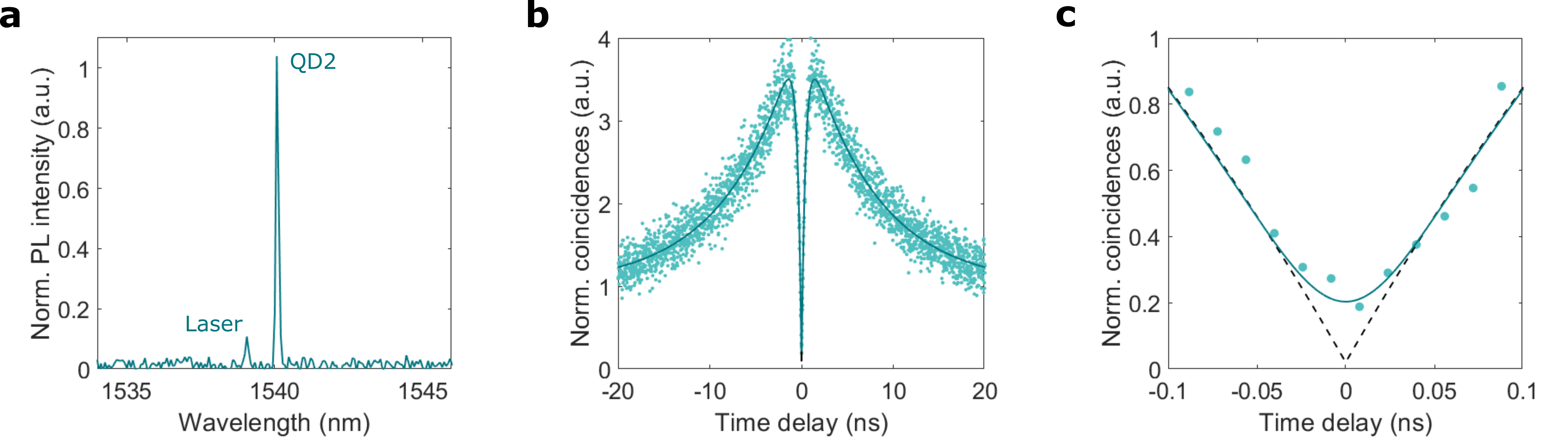}
\caption{\textbf{Device characterization under phonon-assisted excitation.} 
(a) Photoluminescence spectrum of QD2 under phonon-assisted excitation. The small peak at 1539 nm corresponds to residual leakage of the excitation laser through the bandpass filters. 
(b) Second-order correlation function $g^2(\tau)$ of QD2 measured under phonon-assisted excitation. The HBT data is fitted with a three-level function (solid blue line), convolved with a Gaussian function with 70 ps FWHM to account for the detector time resolution. The dashed black line represents the deconvolved fit function.
(c) Zoom-in of the HBT fit around zero delay. The deconvolved fit function (black dashed line) yields $g^2(0) = 0.019(0.049)$.
}
\label{Fig3} 
\end{figure*}

\noindent The circular photonic crystal resonator consists of air holes with diameter \(d\) etched into a semiconductor membrane around a central disk containing the QD emitter. Similar to conventional bullseye cavities, the device also includes an insulating oxide layer and a backside gold mirror  (Figure \ref{Fig1}a). The position of the \(m\)-th air hole in the \(xy\)-plane is given by \cite{Ma.2024}:
\[x = \left[R + \Lambda (N - 1)\right]\cos{\left(\frac{2m\pi}{nN}\right)}\]
\[y = \left[R + \Lambda (N - 1)\right]\sin{\left(\frac{2m\pi}{nN}\right)}\]
\noindent where \(R\) is the radius of the central disk, measured from the device centre to the hole centre in the first ring, \(\Lambda\) is the grating period, \(N\) is the number of rings, and \(n\) denotes the degree of rotational symmetry, which is set to 12 for all devices presented in this work (Figure \ref{Fig1}b).

Numerical finite element method (FEM) simulations indicate that an optimized InP device targeting a resonance wavelength of 1550 nm could achieve Purcell factors \(F_p\)> 20, along with dipole collection efficiencies for a numerical aperture of \(\mathrm{NA} = 0.65\) approaching 90\% (Figure \ref{Fig1}c). As with conventional bullseye designs, the resonance wavelength can be adjusted by varying the radius of the central disk and the periodicity of the grating in the design phase. 
To experimentally validate that a circular photonic crystal resonator can exhibit optical properties comparable to those of conventional CBGs, we fabricate non-deterministic devices on an InP membrane incorporating droplet epitaxy (DE) InAs QDs. By investigating the PL signal under high-power above-band excitation, we observe cavity modes in the telecom C-band. As shown in Figure \ref{Fig1}d, a 10 nm variation in the central disk radius leads to a spectral shift of approximately 7 nm, consistent with the design tunability observed in simulations.
In the low-power PL regime, we identify two devices of particular interest, each hosting a QD spectrally coupled to the fundamental cavity mode (Figure \ref{Fig2}a, see also Supplementary Figure \ref{SupFig_QD_and_modes}). Both display a bright isolated spectral line without any observable fine structure splitting, at wavelengths of 1529 nm (QD1) and 1540 nm (QD2). After spectral filtering, up to 20 million photon counts per second are detected by a superconducting nanowire single-photon detector (SNSPD) when the QD1 transition is pumped with an above-band CW laser (Figure \ref{Fig2}b).

To verify the single-photon nature of the emission, we perform second-order correlation measurements using a fiber-based Hanbury Brown and Twiss (HBT) interferometer. Figure \ref{Fig2}c shows the second-order correlation function \( g^{(2)}(\tau) \), measured at half saturation power. The histogram displays a pronounced antibunching dip at zero delay, and the \( g^{(2)}(0) = 0.107 (0.005) \) extracted by fitting with a 3-level model confirms the high single-photon purity.
Under above-band pulsed excitation at 80~MHz, we measure a photon count rate of 3.35~MHz at saturation. By correcting for the efficiency of the SNSPDs and accounting for the imperfect purity (Supplementary Figure \ref{SupFig_g2_pulsed}), we estimate a 4.5\% end-to-end efficiency of the single-photon source system (consisting of the QD emitter coupled to the resonator, the confocal microscope setup, and the spectral filter).

In Figure \ref{Fig2}d, we present a comparison between the time resolved PL trace of QD1, which exhibits a decay time constant of \( \tau_{\mathrm{QD1}} \sim 800~\mathrm{ps} \), and that of a representative QD located outside of the resonator. To estimate the Purcell enhancement, we extract an average lifetime of \( \tau_{\mathrm{slab}} = 1.91(0.41)~\mathrm{ns} \) from time resolved measurements of ten QD transitions in unpatterned regions of the same sample. Based on this comparison, we determine a Purcell factor of approximately 2. We attribute the modest enhancement primarily to spatial mismatch between the QD emitter and the centre of the resonator.
The second transition of interest exhibits a slightly shorter lifetime \( \tau_{\mathrm{QD2}} \sim 750~\mathrm{ps} \), but a lower single-photon purity. We find that the shape of the \( g^{(2)}(\tau) \) dip is best described by adding a fourth state \cite{Anderson.2020}, resulting in \( g^{(2)}(0) = 0.281(0.006) \) (Figure \ref{Fig2}f).

Different excitation techniques can be employed to enhance the single-photon purity of QD-based sources. Resonant excitation is popular for suppressing background emission and minimizing noise processes associated with off-resonant carrier relaxation pathways \cite{Nawrath.2021, Wells.2023}. While evidence of Rabi oscillations has recently been observed in a QD coupled to a CBG under low pump power \cite{Rickert.2025}, the suppression of reflected laser light remains difficult due to the strong scattering induced by the bullseye pattern. To overcome this challenge, we adopt a phonon-assisted scheme \cite{Quilter.2015, Cosacchi.2019, Thomas.2021}, in which a spectrally detuned laser enables efficient excitation while maintaining spectral separation from the emitted single photons. Figure \ref{Fig3}a shows the PL spectrum of QD2 under phonon-assisted excitation. Compared to the above-band case, this approach leads to a substantial improvement in single-photon purity by over an order of magnitude, yielding \( g^{(2)}(0) = 0.019(0.049) \) (Figures \ref{Fig3}b and \ref{Fig3}c). 
Despite this improvement, we also observe increased blinking on short time scales, as evidenced by the pronounced bunching around the \( g^{(2)}(0) \) dip in Figure \ref{Fig3}b. This phenomenon is typically attributed to the presence of a shelving state that is only accessed through phonon-assisted excitation. Other potential causes may include charging events or spectral wandering due to fluctuations in the QD's charge environment \cite{Vajner.2024}.

%
%
% FIGURE 4
\begin{figure*}[t]
\includegraphics[width=0.67\textwidth]{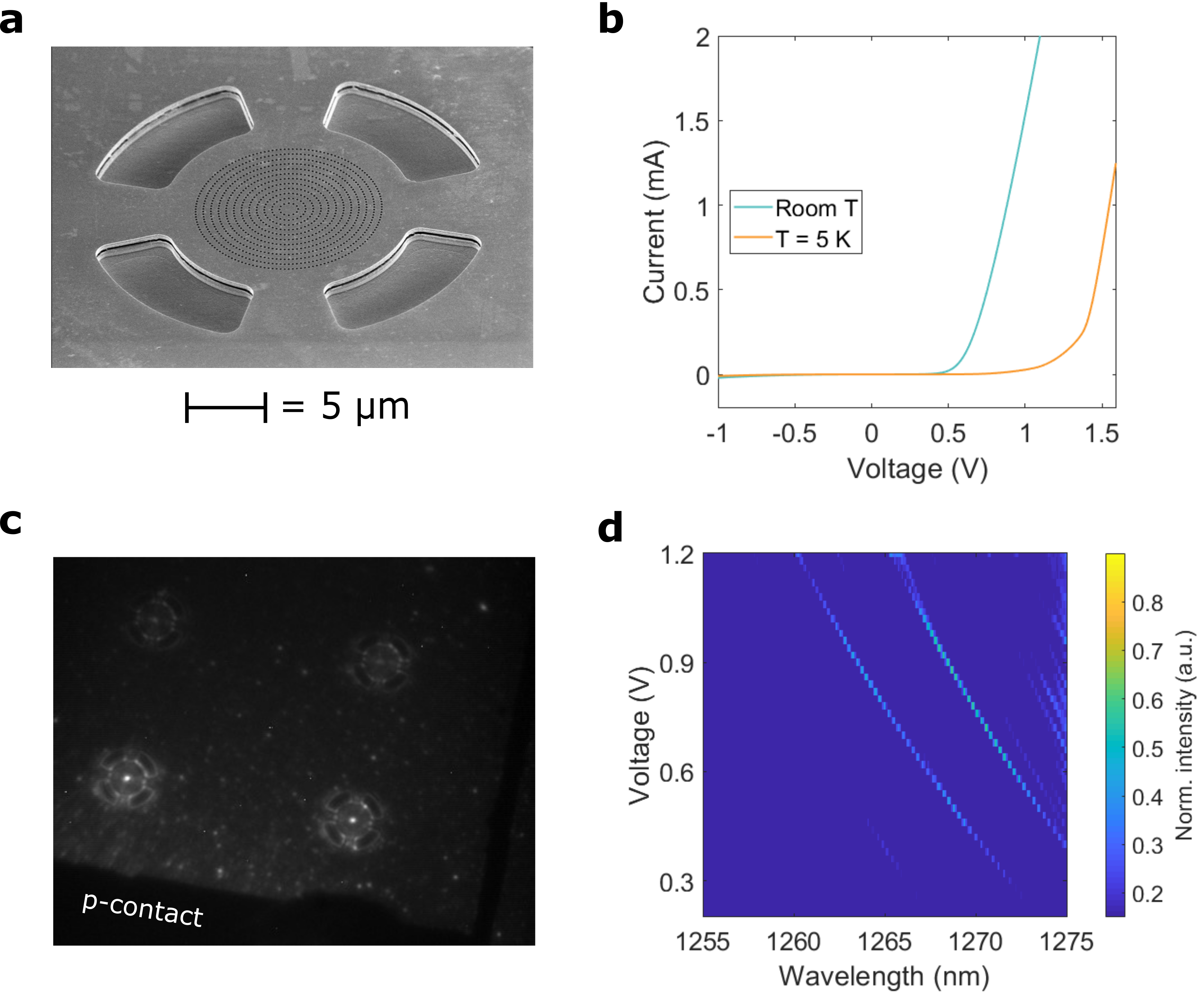}
\caption{\textbf{Electrically contacted resonators.} 
(a) Scanning electron microscope image of an exemplary device fabricated on a suspended GaAs slab. 
(b) I-V characteristic of the diode measured at room temperature and cryogenic temperature.
(c) Microscope image of telecom wavelength emission from QDs coupled to resonators when a forward bias of 1.6 V is applied to the diode. The image is recorded with a 1200 nm long-pass filter in front of the camera to suppress the short wavelength emission.
(d) Photoluminescence spectra recorded on an electrically contacted device under above-band excitation as a function of applied voltage to the diode. A wavelength shift exceeding 10 nm is observed for two isolated QD transitions in the telecom O-band.
}
\label{Fig4} 
\end{figure*}

The results presented thus far demonstrate that C-band QD sources based on circular photonic crystal resonators can achieve performance comparable to conventional CBGs. Building on this, we further demonstrate electric field control of O-band Stranski-Krastanov (SK) QDs embedded in a GaAs slab. These QDs are incorporated into a thin \textit{pin} diode structure, which is more straightforward to implement on the GaAs platform compared to InP, due to the doping challenges associated with the latter. Square mesas with side length of 500 \textmu m are etched into the chip, each containing multiple circular photonic crystal resonators. A sacrificial layer beneath the \textit{pin} slab enables underetching through trenches surrounding each device (Figure \ref{Fig4}a), forming an air gap that creates a weak vertical cavity. Notably, this architecture is also compatible with flip-chip integration of a backside gold mirror for enhanced vertical confinement and photon collection, although the underetch approach requires fewer processing steps. Electrical contact is established via individual \textit{p}-contacts atop each mesa and a shared \textit{n}-contact in the surrounding etched region.

The devices exhibit standard diode I–V characteristics (Figure \ref{Fig4}b), with the turn-on voltage varying with temperature in accordance with the bandgap shift. Under forward bias beyond turn-on, electroluminescence is observed at the resonator center (Figure \ref{Fig4}c), resulting from carrier recombination within the QDs. In addition to enabling direct electrical excitation, the device contacts also allow tuning of optically excited QDs. As shown in Figure \ref{Fig4}d, applying a sub-threshold bias during above-band laser excitation induces a Stark shift of the QD emission wavelength, with a wide range of tuneability exceeding 10 nm. This degree of control is potentially critical for bringing QD transitions into resonance with remote sources \cite{Patel.2010, Pont.2024}, and can also facilitate manipulation of the QD charge state \cite{Wijitpatima.2024}, although this aspect is not explored in the present work. Moreover, the device's compatibility with standard low-voltage power supplies is important for the deployment of sources based on this approach into end-user applications outside a laboratory environment.
%
%%%%%%%%%%%%%%%%%%%%%%%%%%%%%%%%%%%%%%%%%%%%%%%%%%%%%%%%%%%%%%%%%%%%
%% DISCUSSION
%%%%%%%%%%%%%%%%%%%%%%%%%%%%%%%%%%%%%%%%%%%%%%%%%%%%%%%%%%%%%%%%%%%%%

%
\section*{Discussion}
\noindent In summary, we have presented the first experimental realization of optically excited quantum light sources based on InAs/InP QDs coupled to circular photonic crystal resonators. Our devices operate in the telecom C-band and exhibit key figures of merit comparable to those achieved with state-of-the-art InP conventional bullseye resonators \cite{Holewa.2024}, including bright emission and clear antibunching. By employing phonon-assisted excitation, we have further improved the single-photon purity, achieving $g^{(2)}(0) = 0.019(0.049)$.
Additionally, we demonstrated electric field control of telecom O-band InAs/GaAs QDs embedded in electrically contacted devices, achieving a Stark tuning range exceeding 10 nm. 
These results validate the circular photonic crystal resonator as a viable and scalable platform for the development of efficient, tuneable, and electrically controllable quantum light sources at telecom wavelengths.

While the reported results provide a proof-of-principle demonstration, further work is required to make the device performance comparable with the most advanced QD sources and support future integration in quantum photonic networks. In future experiments, deterministic techniques such as in-situ electron beam lithography \cite{Holewa.2024} or cathodoluminescence mapping \cite{Rickert.2025} could be employed to improve the fabrication yield and optimize the spectral and spatial overlap between the target QD transition and the cavity mode. Such approaches would not only enhance control over the emitter–cavity coupling conditions but also lead to improved Purcell enhancement, which is critical for applications in quantum information processing operating at high clock rates.

The broad cavity mode supported by the circular photonic crystal resonator could be used to simultaneously enhance both neutral exciton and biexciton transitions, which are typically separated by a few nanometres in InAs QDs. In this framework, the presence of electrical contacts enables tuning of the fine structure splitting to values below 10 \textmu eV, a key requirement for the generation of high-fidelity entangled photon pairs \cite{Xiang.2020}.
The compatibility of the device with phonon-assisted excitation could also be harnessed to investigate the spin coherence and relaxation dynamics of a single carrier confined in the QD \cite{Coste.2023b}, opening a pathway towards spin–photon interfaces in the telecom range. Finally, elliptical variations of the resonator \cite{Barbiero.2024, Ge.2024} could enable cycling transitions, a crucial ingredient for the generation of photonic cluster states using time-bin encoded qubits \cite{Appel.2021}.
%
 
%

%
%%%%%%%%%%%%%%%%%%%%%%%%%%%%%%%%%%%%%%%%%%%%%%%%%%%%%%%%%%%%%%%%%%%%%
%% METHODS
%%%%%%%%%%%%%%%%%%%%%%%%%%%%%%%%%%%%%%%%%%%%%%%%%%%%%%%%%%%%%%%%%%%%%
\section*{Methods}
\noindent \textbf{Sample structure.} 
The InP QD sample consists of a \textit{n}-doped (100) InP substrate, a 900
nm thick AlGaInAs sacrificial layer, and a 280 nm thick undoped InP
membrane containing DE InAs QDs. After the epitaxial growth, the SiO\textsubscript{2} spacer and Au mirror are integrated in the structure using a wafer-scale membrane transfer technique similar to the one presented in our previous work \cite{Barbiero.2022}. First, a 360 nm thick layer of SiO\textsubscript{2} is deposited on the InP wafer surface, followed by approximately 200 nm of Au. The epitaxial wafer is then flipped and bonded to an InP carrier. Next, the majority of the original InP substrate is reduced to an approximate thickness of 50 \textmu m by commercial mechanical lapping and polishing, while the residual InP substrate and AlGaInAs sacrificial layer are finally removed using a two-step selective wet etch process.

The GaAs QD sample consists of an undoped (100) GaAs substrate, a 200 nm thick Al\textsubscript{0.8}Ga\textsubscript{0.2}As sacrificial layer, and a \textit{pin} GaAs membrane with 50 nm thick doped layers and 140 nm thick intrinsic region. Self-assembled SK InAs QDs are grown in the middle of the intrinsic region.

\noindent \textbf{Device fabrication.} 
For optically pumped devices on InP, a layer of SiN\textsubscript{x} is deposited on the InP surface after the membrane transfer process. The circular holes are defined using electron beam lithography on the SiN\textsubscript{x} layer, which acts as a hard mask. The pattern is transferred to the membrane using fluorine-base dry etch for the SiN\textsubscript{x} hard mask and a chlorine-based dry etch process for InP.

GaAs \textit{pin} diodes with circular photonic crystal resonators are fabricated using a two-step dry etch process. Initially, a pattern of square 500 x 500 \textmu m side length mesas is defined in optical resist, followed by a dry SiCl\textsubscript{4}-based etch to reach the \textit{n}-type layer. AuGeNi is then evaporated to form the \textit{n}-type metallization, and the \textit{n}-type contact is annealed at 420 °C. The top contact is defined on the mesa using optical lithography, with CrAu serving as the \textit{p}-type metallization. In the second etch step, circular resonators are defined using electron beam lithography. The etch depth targets the Al\textsubscript{0.8}Ga\textsubscript{0.2}As sacrificial layer. Finally, after the resist is stripped, the devices are underetched in diluted HCl.

\noindent \textbf{Optical characterization.} 
The devices are operated at a temperature of 5 K in a closed-cycle cryostat. For above-band characterization, the excitation is provided by either a CW or a pulsed laser diode (80 MHz repetition rate) at a wavelength of 785 nm. QD emission from a single device is collected by a fiber-coupled confocal microscope with a NA = 0.5 objective lens. Photons are then sent to a spectrometer equipped with an InGaAs photodiode array for spectral analysis. 
Single QD transitions are isolated using a tunable optical filter with 0.25 nm FWHM bandwidth. After spectral filtering, coincidences are detected using two SNSPDs with 75\% efficiency and 70 ps jitter.

Phonon-assisted excitation is achieved using a linearly polarized CW laser blue-detuned by 1.1 nm from the QD resonance. To isolate the QD emission and suppress residual excitation laser, the collection path incorporates a cascade of four high-extinction band-pass optical filters with 0.8 nm FWHM.

%%%%%%%%%%%%%%%%%%%%%%%%%%%%%%%%%%%%%%%%%%%%%%%%%%%%%%%%%%%%%%%%%%%%%
%% ACKNOWLEDGEMENTS
%%%%%%%%%%%%%%%%%%%%%%%%%%%%%%%%%%%%%%%%%%%%%%%%%%%%%%%%%%%%%%%%%%%%%
\section*{Acknowledgements}
\noindent The authors acknowledge funding from the Ministry of Internal Affairs and Communications, Japan, via the project of ICT priority technology (JPMI00316) ‘Research and Development for Construction of a Global Quantum Cryptography Network’. They further acknowledge funding from QFoundry (project number 48484), which is part-funded by the UK Quantum Technologies Challenge under UK Research and Innovation (UKRI).

\section*{Author contribution}
\noindent T. M\"{u}ller, and R. M. Stevenson, and A. J. Shields guided and supervised the project. L. Ravi grew the InP epitaxial sample under the supervision of J. I. Davies. B. Ramsay grew the GaAs epitaxial sample under the supervision of D. J. P. Ellis. G. Shooter and J. Skiba-Szymanska supported the development of epitaxial samples with experimental characterization. A. Barbiero optimized the device designs using FEM simulations. A. Barbiero and J. Skiba-Szymanska fabricated the devices. A. Barbiero, G. Shooter, and J. Huang performed measurements and analysed data. A. Barbiero wrote the manuscript with contributions from G. Shooter, J. Skiba-Szymanska, and J. Huang. All authors discussed the results and commented on the manuscript. The authors declare that they have no competing financial interests.
%


\bibliography{Hole_CBG_Bibliography}

\newpage
%%%%%%%%%%%%%%%%%%%%%%%%%%%%%%%%%%%%%%%%%%%%%%%%%%%%%%%%%%%%%%%%%%%%%
%% Supplementary
%%%%%%%%%%%%%%%%%%%%%%%%%%%%%%%%%%%%%%%%%%%%%%%%%%%%%%%%%%%%%%%%%%%%%

% Switch to one-column format for supplementary information
\clearpage
\begin{widetext}
\renewcommand{\thesubsection}{\arabic{subsection}}
\renewcommand{\thefigure}{S\arabic{figure}}
\setcounter{figure}{0} % Start figure numbering from S0
\section*{Supplementary Information}
\subsection{Simulations}
To optimize the design parameters of the resonators, we employ 3D FEM simulations to solve the wave equation in the frequency domain. Literature values are used for the refractive indices of all materials. The computational region is enclosed within perfectly matched layers to simulate open boundaries. The QD emission is modeled as an electric point dipole oriented along the y axis and placed in the centre of the device (x=0, y=0) at half of the slab thickness. To calculate the Purcell factor, we monitor the power outflow through the boundaries of the model and normalize it by the power generated by the same dipole source in homogeneous InP.
As shown in Figure \ref{Figure_S1_simulations}a, the electric field profile \(|E^2|\) of the cavity mode is primarily confined in the central disk. Consequently, small adjustments of the resonance wavelength can be achieved by varying the radius \(R\) during the design phase (Figure \ref{Figure_S1_simulations}b).  
\begin{figure*}[h]
\includegraphics[width=0.67\textwidth]{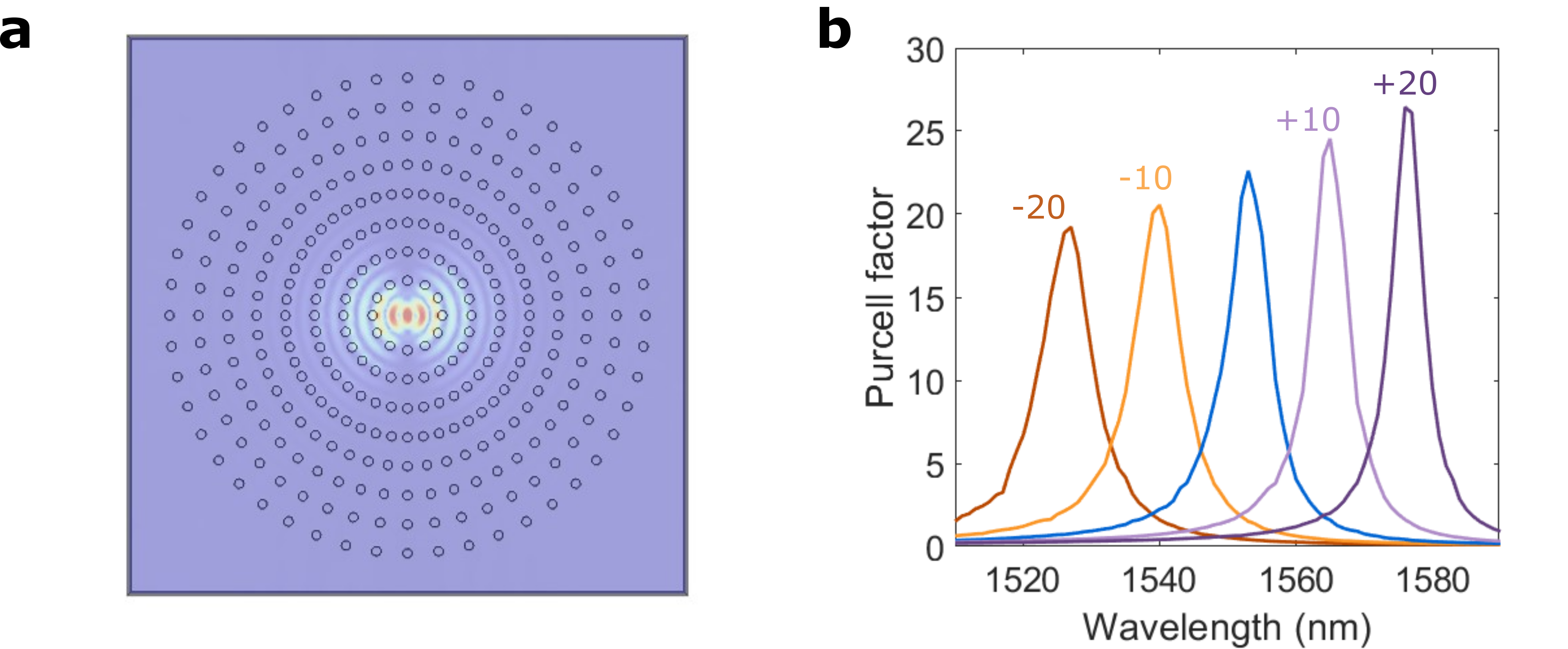}
\caption{\textbf{Simulations.} 
(a) Top view of the simulation domain, showing the electric field profile \(|E^2|\) of the cavity mode. (b) Simulated Purcell factor as a function of wavelength for five InP devices with varying central disk radius. The nominal radius of the central device (blue) is set to 790 nm. Labels indicate the variation in central disk radius measured in nm.
}
\label{Figure_S1_simulations} 
\end{figure*}
\subsection{Extended data}
In addition to the cavity modes measured under high-power above-band excitation presented in Figure \ref{Fig1}d, we report reflectivity spectra from the same devices. Measuring reflectivity is a commonly used alternative method for identifying cavity modes: in this technique, the devices are illuminated using a broadband superluminescent diode, and the reflected signal is collected through the same fiber via an optical circulator. 
Cavity modes appear as dips in the reflectivity spectrum, as shown in Figure \ref{SupFig_cavity_modes}a. We observe an exact correspondence between the reflectivity dips and the PL peaks reported in Figure \ref{Fig1}d, confirming the identification of the cavity modes.
\begin{figure*}[h]
\includegraphics[width=0.67\textwidth]{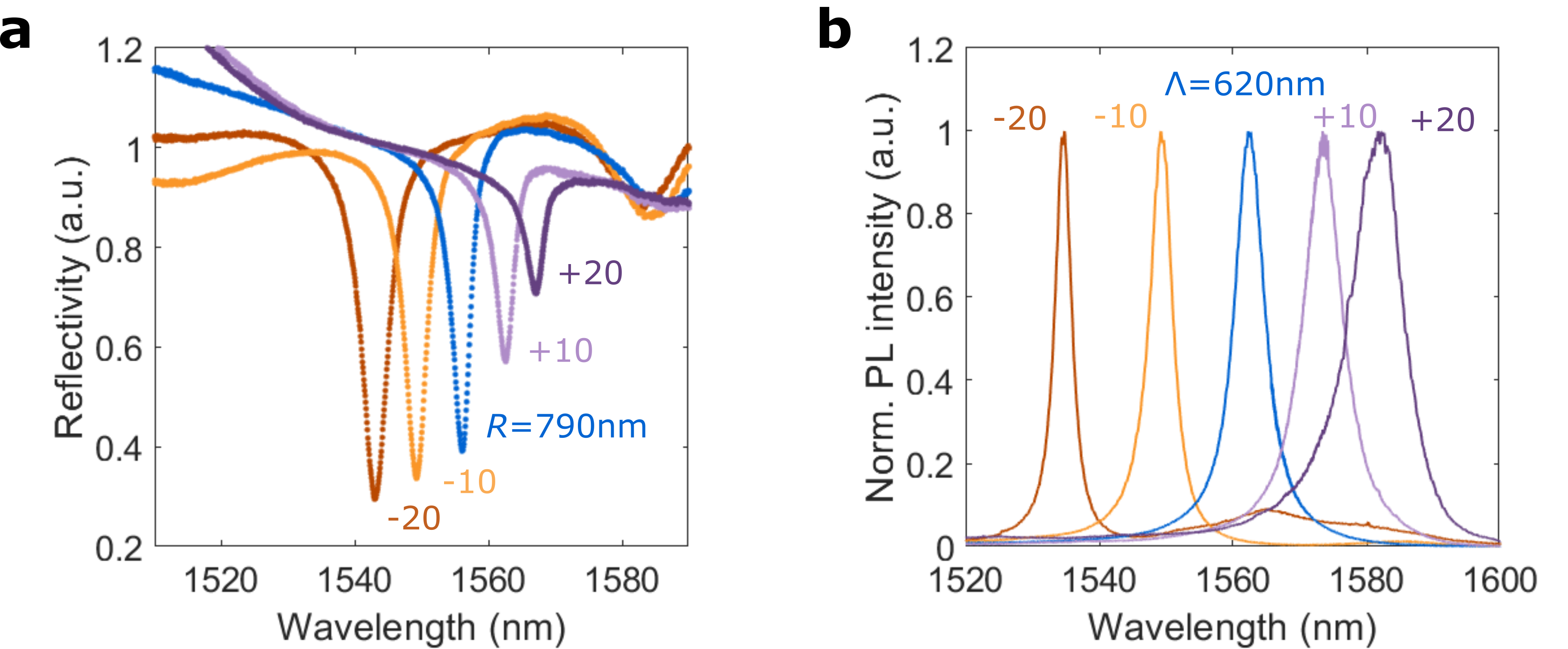}
\caption{\textbf{Cavity modes.} 
(a) Reflectivity spectra measured on  a series of InP devices with different central disk radius. (b) Cavity modes measured under high-power above-band excitation for a series of InP devices with different grating period \(\Lambda\). 
}
\label{SupFig_cavity_modes} 
\end{figure*}
In Figure \ref{SupFig_cavity_modes}b we report an example of cavity modes measured when the grating period \(\Lambda\) is varied. While a regular shift in the resonance wavelength is still observed, notable variations in the cavity quality factor are also evident, as changes in \(\Lambda\) affect the reflectivity property of the grating. 

 Figure \ref{SupFig_QD_and_modes} shows additional PL data measured under CW above-band excitation on the two resonators containing QD1 and QD2. Both devices exhibit a cavity mode in the telecom C-band. The first device (Figure \ref{SupFig_QD_and_modes}a) supports a cavity mode centered at 1526 nm with a FWHM of 5 nm. The QD1 transition is detuned by approximately 3 nm from the center of the mode. The second device (Figure \ref{SupFig_QD_and_modes}b) supports a broader cavity mode with an 8 nm FWHM, showing almost perfect spectral resonance between the QD2 transition and the center of the mode. 
\begin{figure*}[h]
\includegraphics[width=0.67\textwidth]{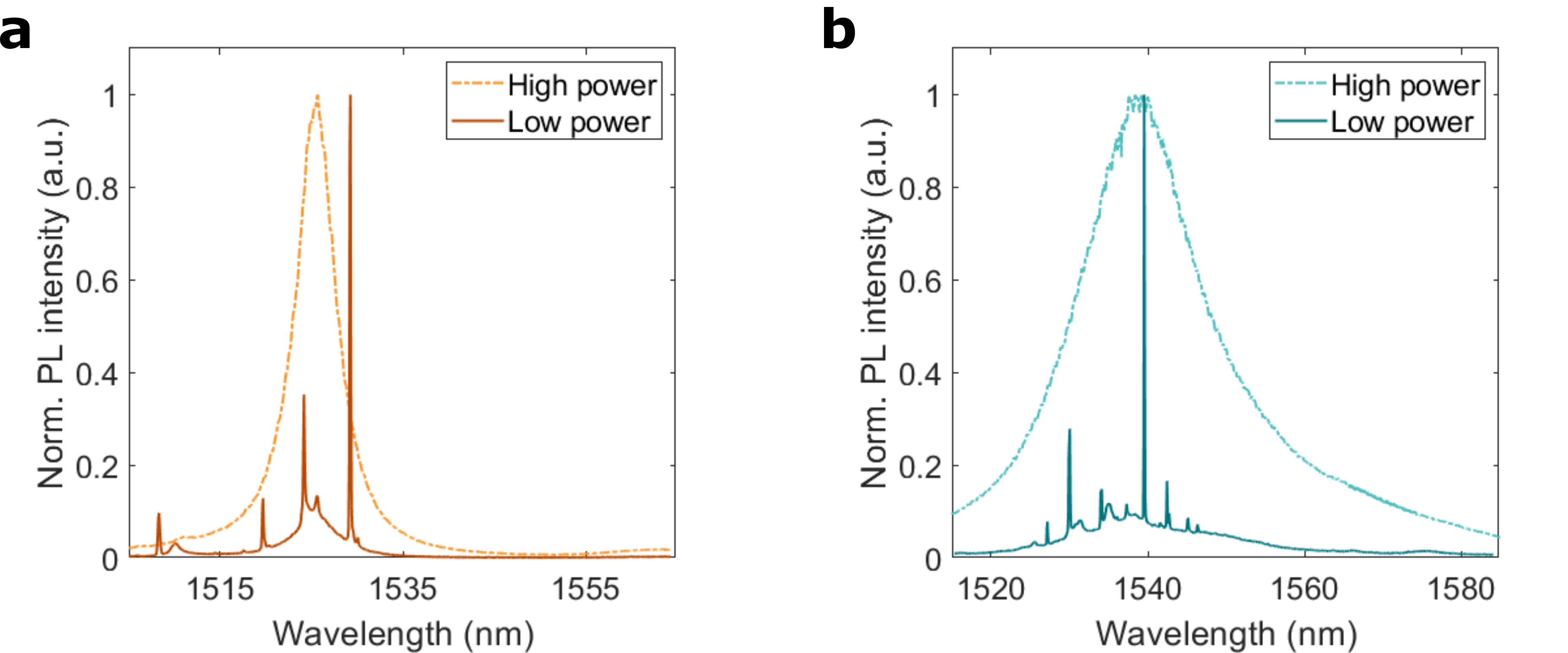}
\caption{\textbf{Photoluminescence spectra.}
PL signal measured under above-band CW laser excitation on the two devices containing (a) QD1 and (b) QD2. The low power spectra show the QD emission, while the high power spectra reveal the fundamental cavity modes supported by the two devices.}
\label{SupFig_QD_and_modes} 
\end{figure*}
%

%\vspace{1em}
Figure \ref{SupFig_g2_pulsed} shows the second-order correlation function \( g^{(2)}(\tau) \) of QD1, measured at saturation under above-band pulsed laser excitation at 80 MHz. The value of the \(g^{(2)}(0) = 0.19\) is used to correct for multiphoton contributions in the calculation of the end-to-end efficiency of the source system.
\begin{figure*}[h]
\includegraphics[width=0.33\textwidth]{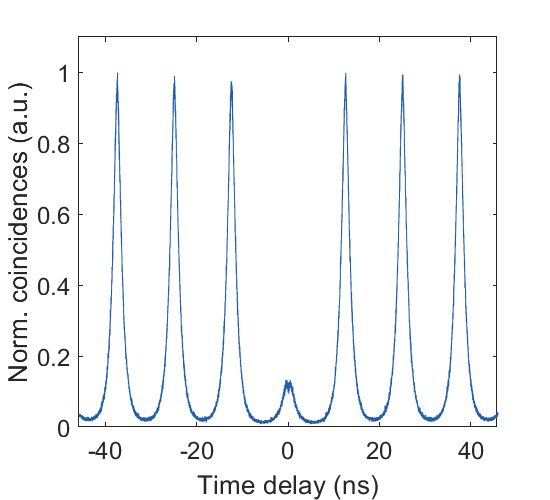}
\caption{\textbf{Pulsed second-order correlation.} 
Second-order correlation histogram of QD1 measured at saturation under above-band pulsed laser excitation at 80 MHz.
}
\label{SupFig_g2_pulsed} 
\end{figure*}
\subsection{Fit of CW second-order correlation data}
The emission of single photons from a QD is often modeled using a simple two-level system, which captures the antibunching dip at the zero time delay. In the above-band CW $g^2(\tau)$ measurements reported in this paper (Figure \ref{SupFig_g2_details}), we observe a small bunching at short delays around $\pm$ 5 ns. This bunching, which we attribute to long-lived shelving states, is captured by adding a third level to the model. 
In the case of QD2 (Figure \ref{SupFig_g2_details}b), we further find that the shape of the antibunching dip is best described by adding a fourth level. As a result, we follow the approach presented in Ref. \cite{Anderson.2020} and define the function:
\[
g_{QD}^{(2)}(\tau )=1-(1+A+B)e^{-|\tau|/\tau_{1}}+Ae^{-|\tau|/\tau_{2}}+Be^{-|\tau|/\tau_{3}},
\]
where $A$ and $B$ quantify the coupling strength from the additional levels. The timescale $\tau_1$ is related to the radiative lifetime of the transition of interest, while $\tau_2$ and $\tau_3$ are the timescales of the additional processes which populate the excited state. It is worth noting that for QD1, $B \rightarrow 0$ and the behavior reduces to a three-level model. Following Ref. \cite{Anderson.2020}, we also define the function:
\[
g_{HBT}^{(2)}(\tau )=\frac{g_{QD}^{(2)}(\tau )+2\beta +\beta^{2}}{(1+\beta)^{2}},
\]
to quantify the influence of uncorrelated background events. Here, $\beta$ is the contribution of the uncorrelated background intensity. 
The function $g_{HBT}^{(2)}(\tau )$ is convolved with a Gaussian function with 70 ps FWHM to account for the detector jitter, and used to fit the correlation data. The values of the $g^2(0)$ reported in the main text are obtained by de-convolving the detector response. The same approach is used to fit the second-order correlation function of QD2 measured under phonon-assisted excitation (Figure \ref{Fig3}b). The quoted errors correspond to one standard deviation resulting from a standard least-squares fitting routine. 
\begin{figure*}[h]
\includegraphics[width=0.67\textwidth]{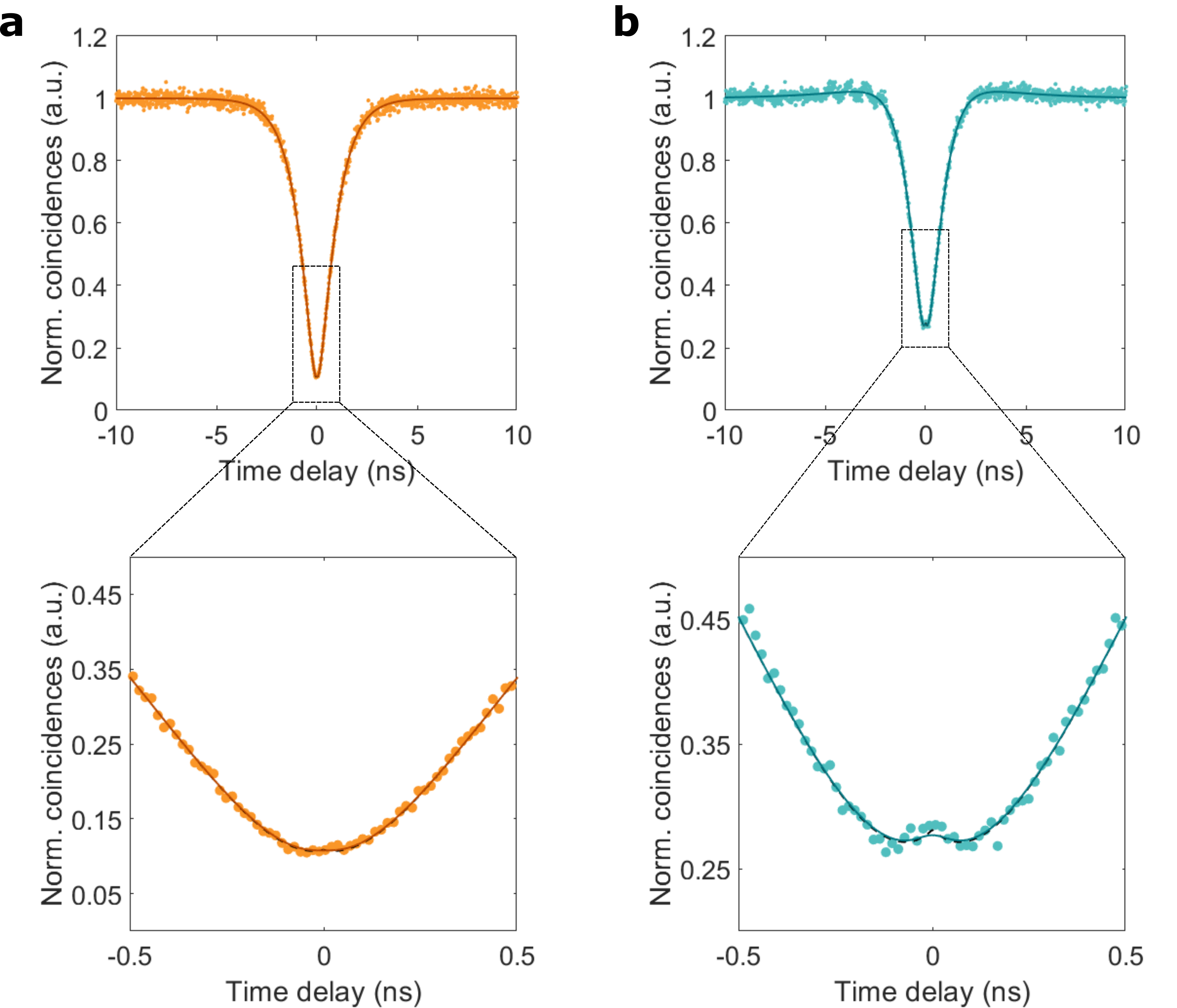}
\caption{\textbf{Second-order correlation fit.} 
Fitted correlation function $g^2(\tau)$ measured on (a) QD1 and (b) QD2 under CW above-band laser excitation at half saturation power. The bottom panels show a detail of the fit on a timescale of $\pm$ 0.5 ns. The solid lines represent the fit of the experimental data, while the dashed black lines represent the de-convolved fit function.
}
\label{SupFig_g2_details} 
\end{figure*}
\end{widetext}
% Return to two-column format
\twocolumngrid

\end{document}